\documentstyle[12pt,psfig]{article}
\begin{document}
\thispagestyle{empty}
\noindent\hbox to\hsize{January 1996 \hfill SNUTP 95-103}\\
\noindent\hbox to\hsize{ revised September 1996 \hfill Brown-HET-1016}\\
\noindent\hbox to\hsize{ hep-ph/9601336\hfill  ULG-PNT-95-1-JRC}\vskip 0.6in
\begin{center}
{\bf BOUNDS ON THE SOFT POMERON INTERCEPT \\ }
\vskip 0.2in
Jean-Ren\'e Cudell\\{\small
Inst. de Physique, U. de Li\`ege,
B\^at. B-5, Sart Tilman, B4000 Li\`ege, Belgium\\
\baselineskip 5pt cudell@gw.unipc.ulg.ac.be}\\
{}~\\
 Kyungsik Kang\footnote{supported in part by the
USDOE contract DE-FG02-91ER 40688-Task A.}
\\
{\small Physics Department, Brown University,
Providence, RI 02912, U.S.A.\\kang@het.brown.edu}\\
{}~\\
and\\~\\
 Sung Ku Kim\footnote{supported in part by the Korea Science and Engineering
Foundation through the Center for Theoretical Physics, Seoul National
University, and by the
Ministry of Education through contract BSRI-95-2427.}\\
{\small Dept. of Physics, Ewha Women's University, Seoul 120-750, Korea\\
skkim@mm.ewha.ac.kr}
\vskip 0.3in
{\bf Abstract}
\end{center}
\begin{quote}
The Donnachie-Landshoff fit of total cross sections has now become a standard
reference for models of total, elastic and diffractive cross-sections. Adopting
their philosophy that simple-pole exchanges should account for all data to
present
energies, we assess the uncertainties on their fits. Our best estimate
for the pomeron intercept is $1.096^{+0.012}_{-0.009}$, but several models
have a good $\chi^2$ for intercepts in the range [1.07,1.11].
\end{quote}\newpage
In Regge theory, the energy dependence of total and elastic cross sections
is implied by the analytic structure of the hadronic amplitude. The
simplest singularities are poles in the $J$-plane, which  correspond to the
existence of bound states of mass $M$ and spin $J$ with
$J=\alpha(M^2)$.
The sum of the exchanges of all the particles on a trajectory $\alpha(M^2)$
leads to a hadronic amplitude which behaves, for $s\rightarrow\infty$ and
$t=M^2<0$ finite, as:
\begin{equation}
{\cal A}(s,t)=C  s^{\alpha(M^2=t)}\pm (s\rightarrow -s)
\end{equation}
The last term, which arises from $s$ to $u$ crossing, is usually included as
a Regge signature factor.  Its sign depends on the charge parity of
the exchange.

Meson trajectories are observed to be linear, as can be understood in a string
model:
\begin{equation}
\alpha(M^2)=\alpha_0 + \alpha' M^2
\end{equation}
with $\alpha_0$ the intercept, and $\alpha'$ the slope of the trajectory.
In particular, total cross sections are related to $Im{\cal A}(s,t=M^2=0)$
by the optical theorem
and involve only the intercept of the trajectory.
Many such trajectories are present, but at high $s$ only
the highest-spin trajectories will contribute: the others are suppressed by a
power of $s$.  Hence, from the observed me\-son spectrum, one expects
total cross section to fall like $1/\sqrt{s}$, as the leading meson
trajectories
are those of the $\rho$,~$\omega$,~$a$ and $f$ mesons,
with $\alpha_0\approx 0.5$.

The fact that total cross sections rise at high energy forces one to introduce
one further ingredient: the pomeron, which in its simplest version
is another Regge trajectory, with an intercept slightly greater than 1.
This object has been hypothesized a long time ago \cite{Collins},
and presumably
arises from the gauge sector of QCD. Its intercept and slope are then
fundamental numbers characterizing the pure gauge sector of Yang-Mills SU(3).
They control both the large-$s$ and the small-$x$ limits
of hadronic cross sections.

There has been a renewed interest in this object
as the range of $s$ and $1/x$ has been expanding, at HERA and at the
Tevatron, where
{\it e.g.} rapidity gaps in deep inelastic scattering have been observed
\cite{gaps}.
The situation at HERA is particularly confusing, as the pomeron intercept
seems to vary with the scale of the problem, and with the masses of external
particles: whereas it seems to be of the order of 1.1 in the total $\gamma p$
cross section, it increases slightly in quasi-elastic production of $\rho$
mesons,
reaches values of the order of 1.25 in $J/\psi$ photoproduction,
and increases further in high-$Q^2$ deep-inelastic
scattering. At the same time, the
gap cross section  remain compatible with a low value of the intercept,
as measured in the pomeron flux.

The only object which is well defined operationally is the soft intercept,
{\it i.e.} the one which controls total cross sections at high-energy. How well
that number is known is thus a central issue for HERA, and will no doubt have
a bearing on the extrapolation of soft cross sections to LHC energies.
Hence we propose in this letter to
re-evaluate the simple-pole parametrisation
and to estimate the errors on the various
parameters.

The information on the soft intercept comes from total cross sections.
As the pomeron controls the high-energy behaviour of the cross section,
it will be most sensitive to the highest energy ({\it i.e.} $pp$ and $\bar p
p$)
data.
There is one caveat to this, as one expects the model to break down at low
energy (because of the presence of sub-leading trajectories) and at
high-energy (because multiple pomeron exchanges must unitarise the cross
sections to make them agree with the Froissart-Martin bound). Hence one
must not only determine the best parameters, but also the range of validity of
the model. One might think of resorting to a $\chi^2$ test in order to do this.
This is not possible directly because of the low quality of the data. Indeed,
the largest compilation of data, made available by the Particle Data Group
\cite{dataset} contains points which are inconsistent at the $16\sigma$ level!
[Note that this point is not shown any longer in the 1996 version of the
particle data book, although it is still present in the compilation of data].

To illustrate these problems and their proposed solution, we first
examine the Donnachie-Landshoff (DL) model in its original form
\cite{DL}. This model fits
the $pp$ and $\bar p p$ cross sections
using a minimal number of trajectories. The higher meson trajectories,
$a/f$ ($C$=+1) and $\rho/\omega$ ($C=-$1), are assumed to be degenerate,
and their contributions are added to a pomeron term in the amplitude.
The data are fitted
for $\sqrt{s}>10$ GeV, as lower trajectories would then contribute
less than 1\%, which
is less than the errors on the data. The result of their fit \cite{DLfit}
is a pomeron
intercept of 1.0808, for which they did not quote a $\chi^2$ or error bars,
but simply mentioned that the $\chi^2$ was very flat near the minimum.

We show in the first column of Table~1 our results for such a fit.
We use the usual definition
\begin{equation}
\chi^2=\sum_i \left({d_i-s_i^{-1}  Im{\cal A}(s_i,0))\over e_i}\right)^2
\label{chi2}
\end{equation}
with $d_i\pm e_i$ the measured  $pp$ or $\bar p p$ total cross section
at energy $\sqrt{s_i}$, and
\begin{equation}
Im{\cal A}(s,0)=C_- s^{\alpha_m} + C_+ s^{\alpha_m} +C_P s^{\alpha_P}
\label{amplitude}
\end{equation}
where $C_-$ flips sign when going from $pp$ to $\bar p p$. We shall use the
notation $\alpha_P=1+\epsilon$ in the following.
\begin{center}
{}~\vglue-12pt
{\small \begin{tabular}{|c|c|c|c|}\hline
parameter&  all data&filtered data (2$\sigma$)&filtered data
(1$\sigma$)\\\hline
$\chi^2	$	&410.8	&80.3 &32.4\\
$\chi^2$ per d.o.f.	&3.16 &0.62&0.25\\\hline
pomeron intercept-1
&$0.0912^{+0.0077}_{-0.0070}$&$0.0887^{+0.0079}_{-0.0071}$
&$0.0863^{+0.0096}_{-0.0084}$\\
pomeron coupling (mb)&$19.3^{+1.5}_{-1.7}$& $19.8^{+1.6}_{-1.7}$ &
$20.4^{+1.8}_{-2.1}$ \\\hline
$\rho/\omega/a/f$ intercept-1
&$-0.382^{+0.065}_{-0.071}$&$-0.373^{+0.067}_{-0.073}$
&$-0.398^{+0.083}_{-0.090}$\\
$\rho$/$\omega$ coupling $C_-(pp)$ (mb)
&$-13.2^{+4.1}_{-6.5}$&$-13.3^{+4.3}_{-6.9}$&$-15.3^{+5.7}_{-10.1}$\\
$a/f$ coupling (mb)		&
$69^{+20}_{-13}$&$62^{+19}_{-12}$&$67^{+25}_{-16}$\\\hline
\end{tabular}}~\\
\end{center}
\begin{quote}
{\small Table 1: Simple pole fits to total $pp$ and $\bar p p$ cross sections,
assuming
degenerate $C=+1$ and $C=-1$ exchanges. }
\end{quote}

The value of the $\chi^2=410$ for 135 data points and 5 parameters is
totally unacceptable, as it corresponds to a confidence level (CL) of
$2\times 10^{-36}$! This CL gives the probability that the model with fixed
parameters could generate the data through a random fluctuation.
Note that all other fits have so far suffered
from the same problem \cite{Kangfit}.
There are two possible outcomes to such a high value of the $\chi^2$:
either the model is to
be rejected, or some of the data are wrong.
As we already mentioned, there are a few
obviously wrong points within the data. Hence before reaching conclusions
about the model, one needs to tackle the issue of eliminating those points.

We propose here a reasonable criterion, which will give us the order of
magnitude of the uncertainties.
The best selection criterion would based on physics arguments to decide
which experiments got the wrong results.
Unfortunately, besides experimental questions, this would involve some
personal bias, and prove infeasible for old (ISR) data, where most of the
incompatibilities lie. Here, we propose the following approach, which is
independent of any underlying theoretical model. Most data are bunched
in several energy intervals. We ask that a given data point should not deviate
by more than 1 or 2$\sigma$
from the average of all data in a bin of $\pm 1$ GeV centered around it
(note that filtering at the 1$\sigma$ level rejects both  the E710 and the
CDF measurements, whereas both are kept when filtering at the 2$\sigma$ level
).
All the data which do not fulfill this criterion are to be rejected. This
eliminates some of the pathological points in a fit-independent manner,
and brings in a dramatic improvement in the  $\chi^2/dof.$, which falls
down to less than 1 for the best fits.
Note that this procedure is not unlike the one
followed by UA4/2 in \cite{UA4}.
The number of standard deviations, as well as the width of the
intervals considered are of course somewhat arbitrary. Because the data
are concentrated at certain energies, the width does not matter too much,
but one could certainly change the filtering to an arbitrary number of
standard deviations. When considering large numbers of points, it is in fact
quite likely that about one third will deviate by more than one standard
deviation. Hence the $1\sigma$ filtering is presumably too stringent,
although we
shall see that the results are stable when going from one selection
criterion to the other.

We give in Table~2 the number of points kept when filtering the data,
the full data sets being available at
{\small http://nuclth02.phys.ulg.ac.be/DATA.html}. Note that in the following
we shall also consider an alternative data set \cite{Kangdata}. Its main
difference with that of the Particle Data Group is that the statistical
and systematic errors have been added in quadrature, and that it includes
the measurements of the $\rho$ parameter.
\begin{quote}{\small
\begin{tabular}{|c|c|c|c|c|}\hline
data set&$\sigma_{tot}^{pp}$ (mb) &$\sigma_{tot}^{\bar pp}$ (mb)&$\rho^{pp}$&$
\rho^{\bar pp}$ \\ \hline
P.D.G. \cite{dataset}&94&28&-&-\\
P.D.G. \cite{dataset} - $2\sigma$ &84&28&-&-\\
P.D.G. \cite{dataset} - $1\sigma$ &65&20&-&-\\
Ref.~\cite{Kangdata}&66&29&41&13\\
Ref.~\cite{Kangdata} - $2\sigma$&60&28&38&13\\
Ref.~\cite{Kangdata}- $1\sigma$&53&19&31&13\\\hline
\end{tabular}}
{}~\\
\end{quote}\begin{quote}
{\small Table 2: The number of points kept after data selection, for \\
\hbox{$\sqrt{s}>10$~GeV.}}
\end{quote}

If we now go back to the first line of Table~1, we see that the elimination of
a handful of points drastically changes the $\chi^2$, and makes the model
entirely acceptable. We can now proceed to evaluate the errors on the N
parameters \cite{PDG,minuit,statistics}.
Here the data are given, and we are asking what is the effect
of a random change of parameters. The value of $\chi^2-\chi^2_{min}$ is
distributed as a $\chi^2$ distribution with N parameters. We choose here
to quote the intervals corresponding to the projection of the constant $\chi^2$
hypersurface containing 70\% of the probability. This corresponds to $\chi^2=
\chi^2_{min}+6.06$ in the DL case with 5 parameters. Note that we do not
trust the validity of the method used by the Particle Data Group in their
fit to the DL model, as they simply renormalise the $\chi^2$ to $\chi^2/dof.=1$
and let the new $\chi^2$ vary by one unit.

An important test of the data filtering method is that the central values
and their errors should not depend too much on the filtering itself.
Table~2 shows that this is indeed the case for our fit. We see
that the pomeron intercept is determined to be about 1.090, and that
it could be as high as 1.096. Filtering the data does change the value of
$\chi^2_{min}$, but does not affect its variation around the minimum too much.
Hence the stability of the parameter values seems a more
reasonable criterion than the value of $\chi^2_{min}$.

We can now tackle the question of the energy range of validity of the model.
The two basic requirements are that the $\chi^2/dof.$ be of the order of 1,
and that the determination of the intercept be stable. We show in Fig.~1
the result of varying the energy range. Clearly, the lower trajectories seem
to matter for $\sqrt{s}_{min}<10$ GeV, whereas the upper energy does not seem
to
modify the results (in other words, there is no sign of the onset of
unitarisation). Hence
we adopt $\sqrt{s}_{min}=10$ GeV as the lowest energy at which the model is
correct. This happens to be the point at which $\chi^2_{min}$ is lowest, too.
This dependence on the lower energy cut explains why both
the Particle Data Group \cite{PDG} and Bueno and Velasco \cite{BV}
obtain an wrong value for the intercept, much lower than ours.

{}~\\
\hbox{
\psfig{figure=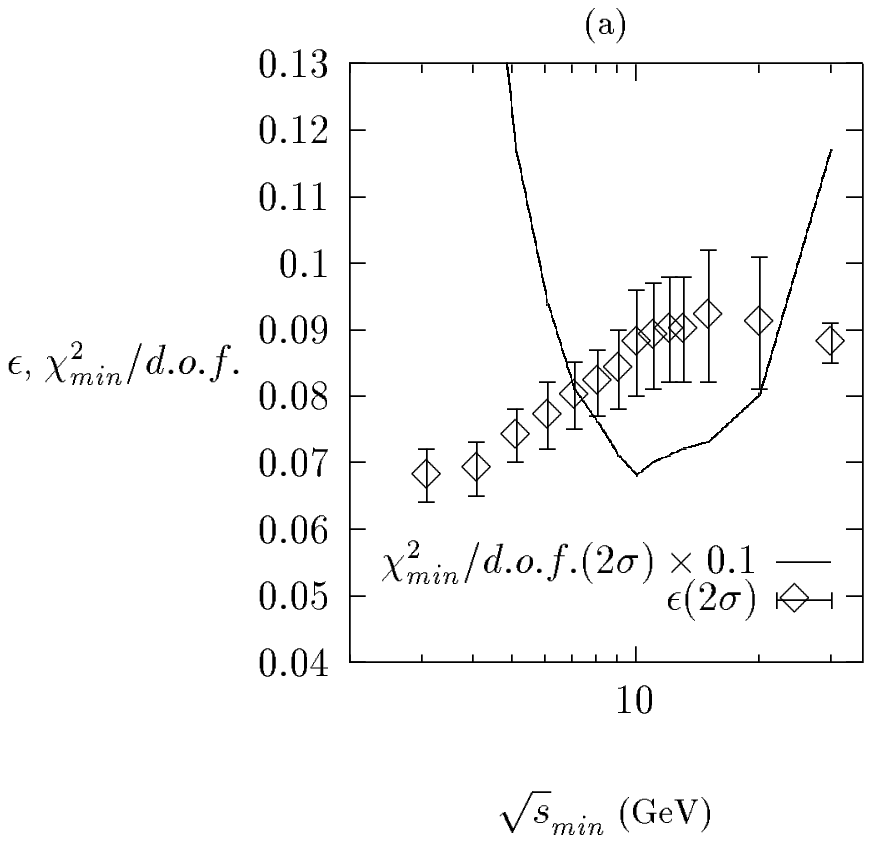,bbllx=2.5cm,bblly=15cm,bburx=12cm,bbury=23cm,width=6cm}
\psfig{figure=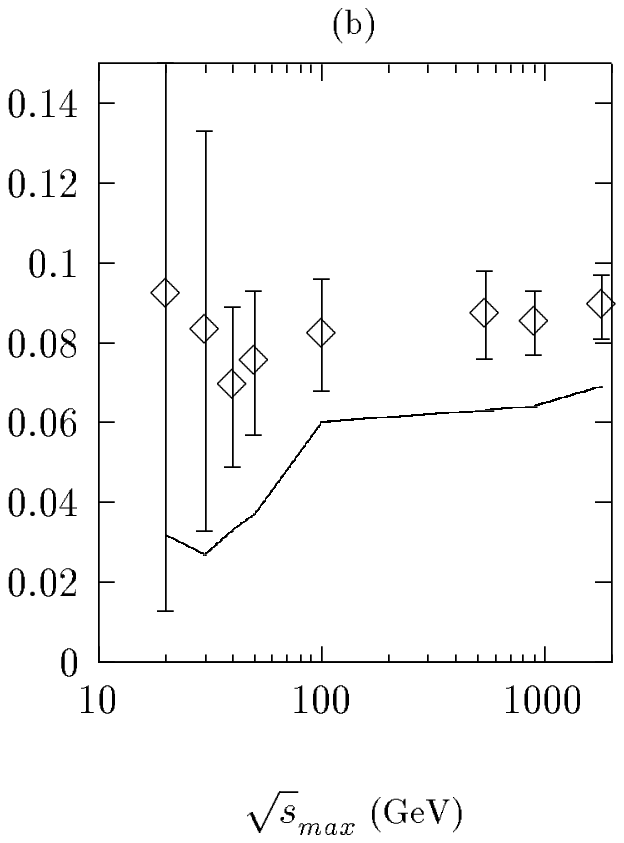,bbllx=2.5cm,bblly=15cm,bburx=12cm,bbury=23cm,width=6cm}
}
\begin{quote}
{\small Figure 1: DL intercept-1 as a function of the lower (a) and upper (b)
energy cuts
on the data. The curve shows the $\chi^2/dof.$ for data filtered at
the $2\sigma$ level.
}
\end{quote}

One must wonder if it is possible to get a better determination of the soft
pomeron intercept by using more data. After performing their fit on
$pp$ and $\bar pp$ cross section, DL extended it to include all measured
hadronic total cross sections. They found that this did not affect the
value of the intercept very much. This is due to the fact that the
other total cross sections have large errors, and that new parameters
are introduced (namely the couplings of the Regge trajectories) each time
one considers a new type of cross section.
The Particle Data Group (PDG) \cite{PDG} obtained very narrow determinations
of the pomeron intercept from the other hadronic reactions. We believe
that their conclusions are wrong, and illustrate this in the case of
the $\pi^{\pm} p$ total cross sections, for which they use
$\sqrt{s}_{min}\approx 4$ GeV and obtain an intercept of $1.079\pm 0.003$.

{}~\\
\hbox{
\psfig{figure=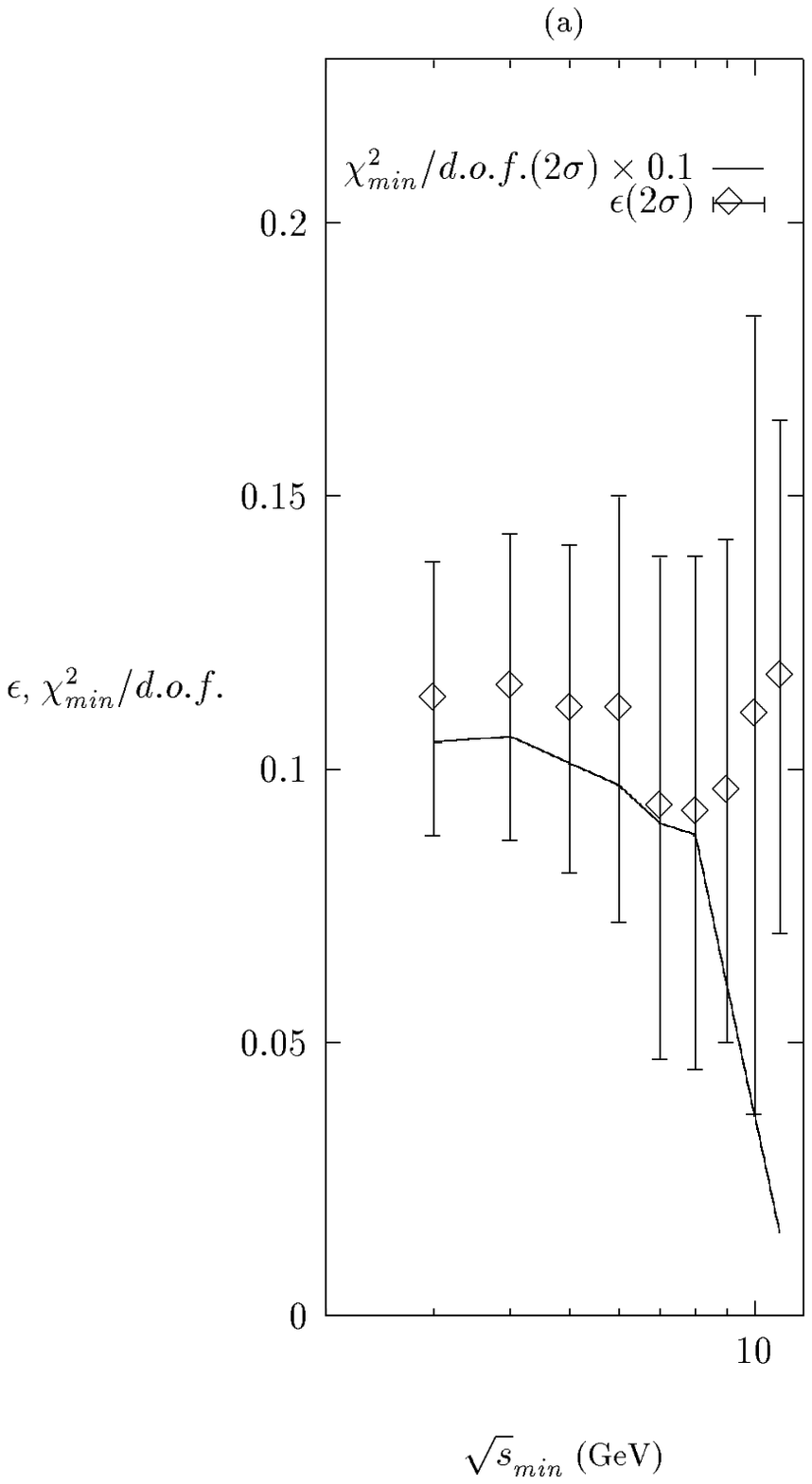,bbllx=2.5cm,bblly=7.5cm,bburx=12cm,bbury=23cm,width=6cm}
\psfig{figure=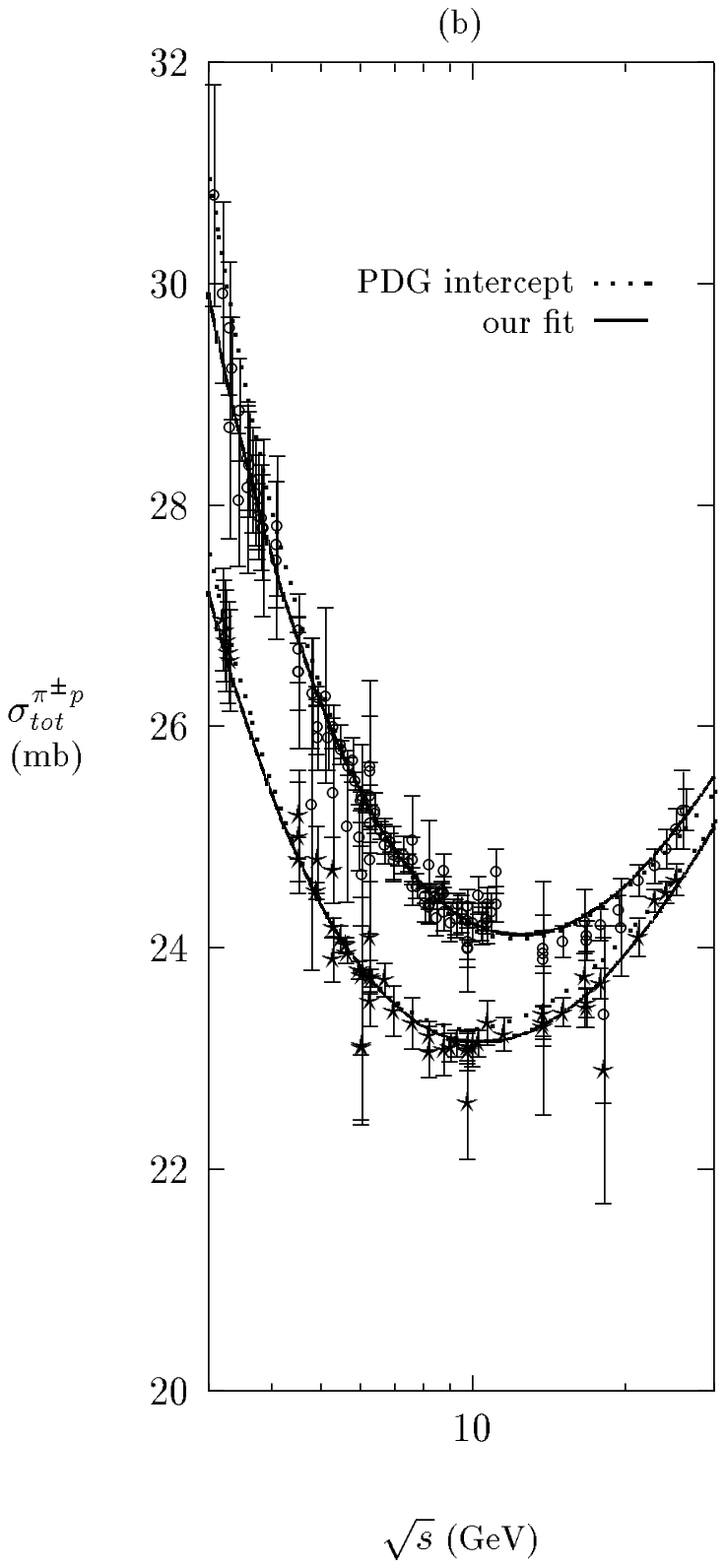,bbllx=2.5cm,bblly=7.5cm,bburx=12cm,bbury=23cm,width=6cm}
}
\begin{quote}
{\small Figure 2: Pomeron intercept from $\pi p$ data: (a) shows the values of
the
intercept as the lower energy cut on the data is changed; (b) shows our
best fit together with that of the Particle Data
Group, together with the set of data filtered at the $2\sigma$ level.
}
\label{pirange}
\end{quote}

We show in Fig.~2 our results for such a fit: for data filtered at the
$2\sigma$ level (92 points), and for $\sqrt{s}_{min}=4$ GeV, we obtain
$\alpha_0=1.115^{+0.030}_{-0.023}$. There is no significant change if
we modify the lower energy cutoff on the data. We show in Fig.~(2.b) our
best fit together with that of the Particle Data Group. Although according
to their estimate our central
value for the intercept is 10 standard deviations from theirs,
we see
that the two fits are indistinguishable. Hence we believe that both their
standard deviations and their central values are wrong. These conclusions are
not affected by the use of the full data set instead of the one filtered at
the $2\sigma$ level. The above value of the intercept in fact gives a slightly
smaller $\chi^2/dof$ than the one quoted in the Particle Data Book.
Note that this intercept is consistent with the one we got
from the analysis of $pp$ and $\bar p p$ total cross sections.
The conclusion from this exercise is that the errors from the low-energy
hadronic data are large, especially if we use a low-energy cut-off of the
order of 10. Hence we want to limit ourselves to
$pp$ and $\bar p p$ amplitudes at $t=0$.

\begin{center}
{\small \begin{tabular}{|c|c|c|c|}\hline
parameter&  all data& filtered data ($2 \sigma$)&filtered data ($1\sigma$)\\
\hline
$\chi^2	$		& 561.3	&168.3 &94.9\\
$\chi^2$ per d.o.f.	& 3.28	&1.07&0.77\\\hline
pomeron intercept--1	&$0.0840\pm
0.0050$&$0.0817^{+0.0055}_{-0.0053}$&$0.0804^{+0.0064}_{-0.0061}$\\
pomeron coupling (mb)	& $20.8 \pm 1.1$&$21.4\pm 1.1$& $21.8\pm 1.3$\\\hline
$\rho/\omega/a/f$
intercept--1&$-0.408^{+0.032}_{-0.033}$&$-0.421^{+0.034}_{-0.036}$&
$-0.431^{+0.037}_{-0.040}$\\
$\rho$/$\omega$ coupling $C_-(pp)$ (mb)&$-14.0 ^{+2.6}_{-3.3}$&$-16.5
^{+3.3}_{-4.2}$&$-17.7^{+3.7}_{-4.9}$\\
$a/f$ coupling (mb)		& $67.0 ^{+7.6}_{-6.7}$&$66.6 ^{+8.3}_{-7.2}$&$67.6
^{+9.0}_{-7.8}$\\\hline
\end{tabular}}~\\
\end{center}\begin{quote}
{\small Table 3: Simple pole fit to total $pp$ and $\bar p p$ cross sections,
and to the $\rho$
parameter, assuming
degenerate $C=+1$ and $C=-1$ meson exchanges.}
\end{quote}

One more
piece of data can be used however:
the knowledge of the intercept is sufficient to determine the value
of the real part of the amplitude, using crossing symmetry, and hence
the measurements of the $\rho$ parameter provide an extra constraint.
We use the data
collected in Ref.~\cite{Kangdata}, and obtain a somewhat worse fit,
shown in Table~3, even
when filtering data at the $1$ or $2\sigma$ level. For the 2$\sigma$ filtering
of the data, the confidence level goes from 99.4\% to 36\%.
We show the curves corresponding to the second column of Table~3 in Fig.~3 with
dotted lines.
Whether one
should worry about this, and about the change of central value for the
parameters, is a matter of taste.
Note however that it is this small change of
central values, combined with the effect of too low an energy cut, that lead
Bueno and Velasco \cite{BV} to conclude that simple-pole parametrisations were
disfavored.

{}~\\
\hbox{
\psfig{figure=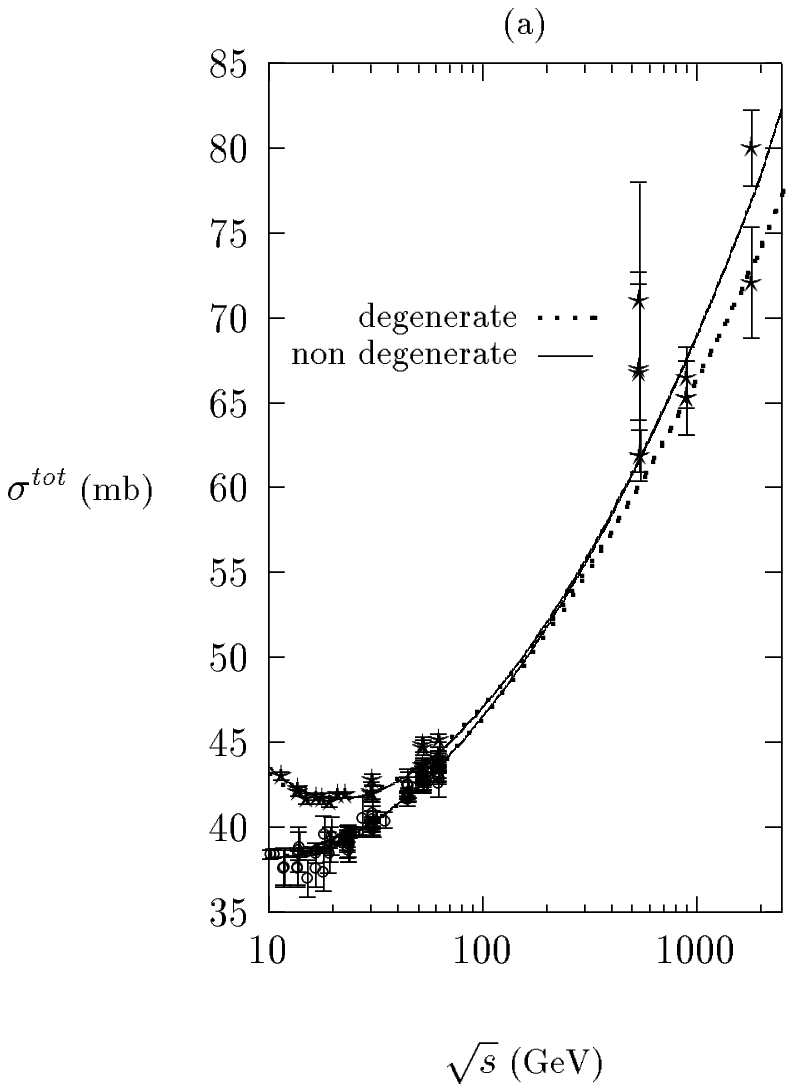,bbllx=2.5cm,bblly=12cm,bburx=12cm,bbury=23cm,width=6cm}
\psfig{figure=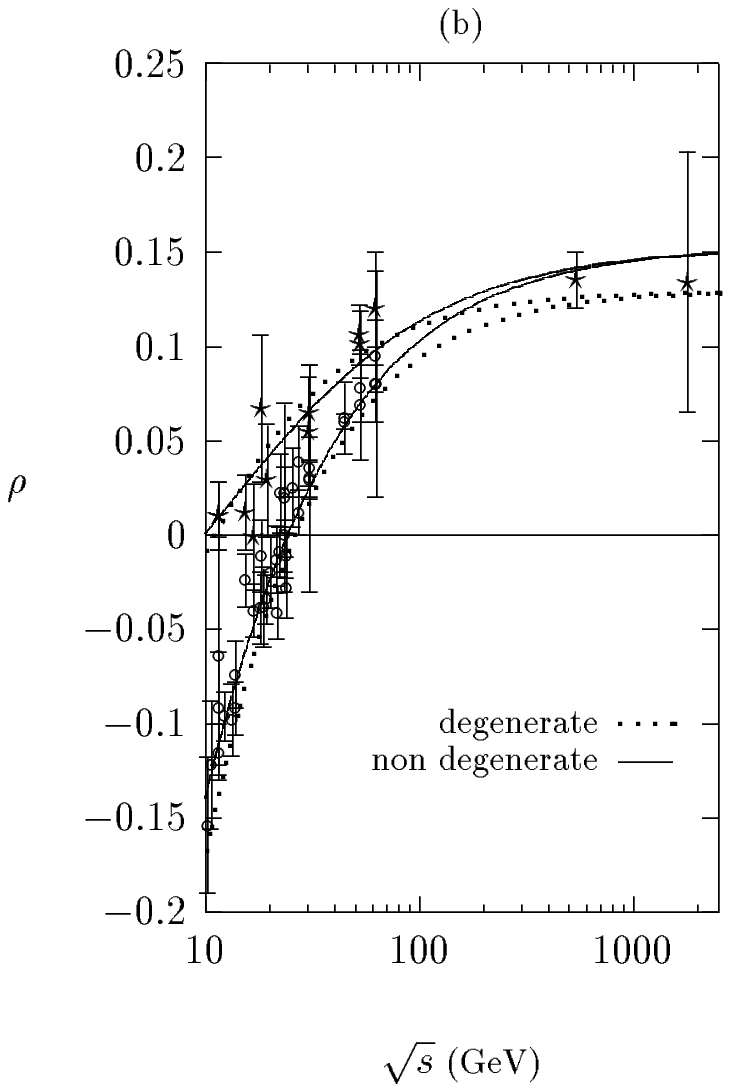,bbllx=2.5cm,bblly=12cm,bburx=12cm,bbury=23cm,width=6cm}
}
\begin{quote}
{\small Figure 3:  Best fits to $2\sigma$ filtered data. The dotted lines
correspond to
the original DL model given in Eq.~(\protect{\ref{amplitude}}), whereas the
plain ones correspond
to a model where the degeneracy of the lower trajectories is lifted, as in
Eq.~(\protect{\ref{2traj}}). The data points are the PDG data filtered at
the $2\sigma$ level.
}
\label{bestfits}
\end{quote}

Before concluding on the best value of the intercept, we need to examine
the influence of low energy physics on the determination of the intercept.
Although the energy cut $\sqrt{s}_{min}$ eliminates sub-leading meson
trajectories, there is still an ambiguity in the treatment of the leading
meson trajectories. In fact, we shall now see that a slightly different
treatment to that of DL leads to a better $\chi^2$ and to more
stable parameters.
Indeed, there is no reason to assume that the $\rho$, $\omega$, $f$ and
$a$ trajectories are degenerate. We show the data for the
meson mass spectrum \cite{PDG}
 in Fig.~4, as well as the separate best fits.

{}~\\
\centerline{
\psfig{figure=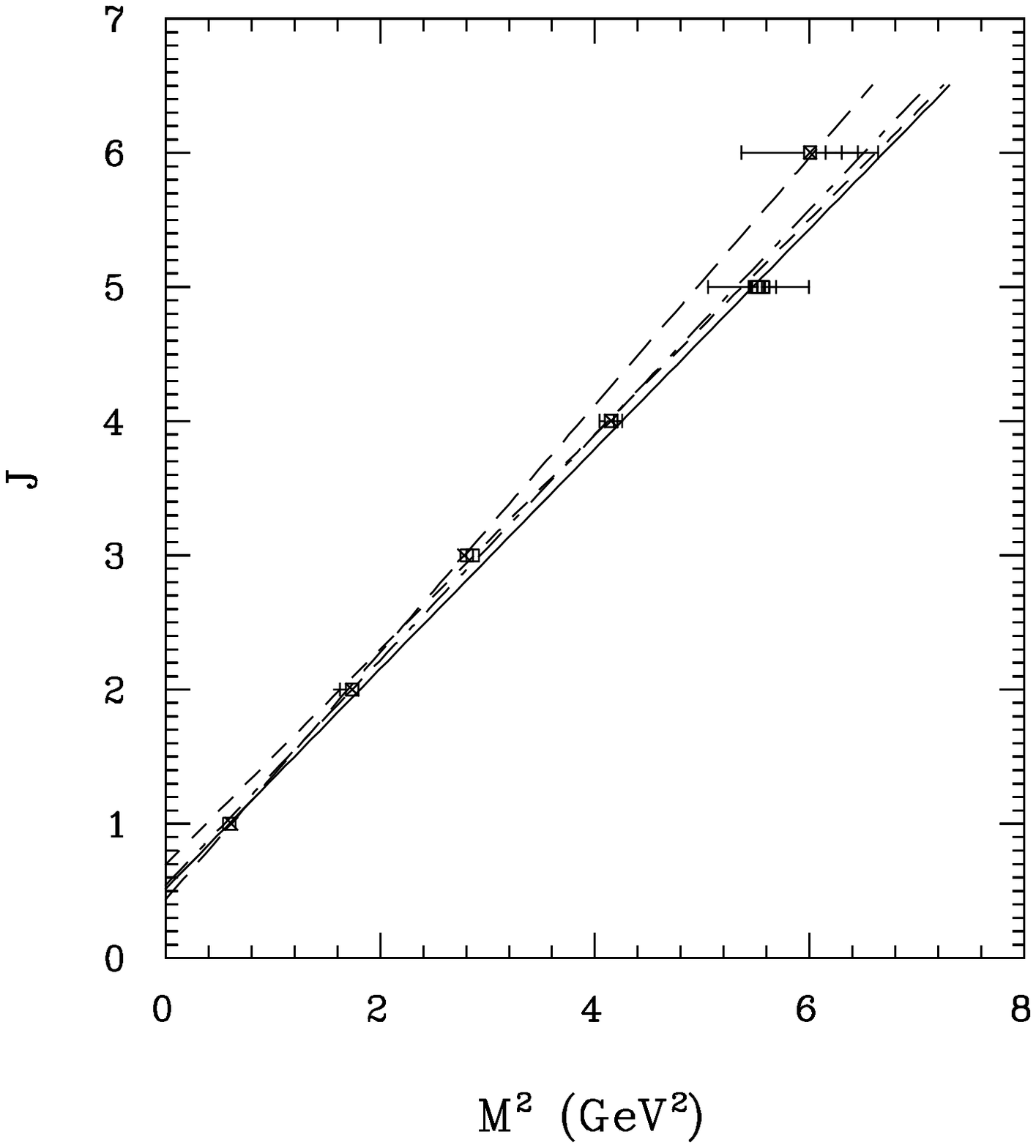,bbllx=2.5cm,bblly=9cm,bburx=17cm,bbury=23cm,width=5cm}
}~\\~\\
\begin{center}
{\small Figure 4: Best fit to the lower trajectories.}
\label{traj}
\end{center}

The trajectories plotted correspond to the following equations:
\begin{eqnarray}
    \alpha_\omega(t)&=& (0.3467\pm 0.0043)+(0.9213\pm 0.0071) t\nonumber\\
   \alpha_\rho(t)&=& (0.5154\pm 0.0014)+ (0.8198\pm 0.0014)t\nonumber\\
    \alpha_a(t)&=&(0.541\pm 0.061)+( 0.839\pm  0.035) t\nonumber\\
   \alpha_f(t)&=& ( 0.697\pm 0.021)+( 0.801 \pm  0.012) t\label{trajec}
\end{eqnarray}
At this point, it is tempting to use the central values of Eq.~(\ref{trajec})
for the lower trajectories. Such a procedure produces a
a pomeron intercept of 1.095 with acceptable $\chi^2$. However, such
a procedure is not satisfactory because it excludes the possibility of a
mixing of the $f$ and the pomeron, or of the simultaneous exchange
of several trajectories.

On the other hand, the data are not constraining enough to determine the
effective intercepts of the four
meson trajectories together with the pomeron intercept. We adopt
an intermediate approach, which is to assume the exchange of separate + and $-$
trajectories with independent intercepts:
\begin{equation}
Im{\cal A}(s,0)=C_- s^{\alpha_-} + C_+ s^{\alpha_+} +C_P s^{\alpha_P}
\label{2traj}
\end{equation}
The resulting numbers are shown in Table~4, and are plotted in
Fig.~3 with plain lines.  The $\chi^2$ is smaller than previously,
and the parameters are more stable. The bounds on the soft pomeron intercept
hardly depend on the criterion used to filter the data, and
intercepts as large as 1.108 are allowed.
\begin{center}
{\small \begin{tabular}{|c|c|c|c|}\hline
parameter& all data &filtered data ($2\sigma$)&filtered data ($1\sigma$)\\
\hline
$\chi^2	$		& 505.4&119.6&57.6\\
$\chi^2$ per d.o.f.	& 2.99		& 0.77&0.47\\\hline
pomeron intercept--1	&$0.0990^{+0.0099}_{-0.0088}$&
$0.0964^{+0.0115}_{-0.0091}$&$0.095^{-0.013}_{+0.010}$\\
pomeron coupling (mb)	&$17.5^{+1.9}_{-2.0}$&
$18.0^{+2.0}_{-2.2}$&$18.2^{+2.3}_{-2.6}$\\\hline
$\rho/\omega$ intercept--1	&
$-0.494^{+0.056}_{-0.066}$&$-0.498^{+0.057}_{-0.067}$
&$-0.510^{+0.064}_{-0.077}$\\
$\rho$/$\omega$ coupling $C_-(pp)$ (mb)& $-24.0^{+6.8}_{-10.9}$ &
$-26.5^{+7.7}_{-12.5}$&$-28.2^{+8.9}_{-15.4}$\\\hline
$a/f$ intercept--1		&$-0.312^{+0.051}_{-0.052}$&$-0.315\pm 0.058$&$-0.324\pm
0.066$\\
$a/f$ coupling (mb)		&
$56.8^{+8.1}_{-6.7}$&$54.9^{+9.0}_{-7.2}$&$56.2^{+9.9}_{-7.8}$\\\hline
$\sigma_{tot}(1.8$ TeV$)$ (mb)	&
$77.6^{+2.5}_{-2.7}$&$76.8^{+2.9}_{-2.7}$&$76.4^{+3.4}_{-3.1}$\\\hline
$\sigma_{tot}(10$ TeV$)$ (mb)	&
$108.4^{+7.0}_{-6.7}$&$106.4^{+7.9}_{-6.7}$&$105.4^{+9.2}_{-7.5}$\\\hline
$\sigma_{tot}(14$ TeV$)$ (mb)	&
$115.8^{+8.3}_{-7.7}$&$113.5^{+9.3}_{-7.8}$&$112.3^{+10.8}_{-8.6}$\\\hline
\end{tabular}}~\\
\end{center}\begin{quote}
{\small Table 4: Simple pole fit to total $pp$ and $\bar p p$ cross sections,
and to the $\rho$
parameter, with non-degenerate $C=+1$ and $C=-1$ meson exchanges.}
\end{quote}

\begin{center}
{\small \begin{tabular}{|c|c|c|c|}\hline
parameter& all data &filtered data ($2\sigma$)&filtered data ($1\sigma$)\\
\hline
$\chi^2	$		& 197.9&107.7&56.3\\
$\chi^2$ per d.o.f.	& 1.39		& 0.82&0.52\\\hline
pomeron intercept--1	&$0.0955^{+0.0097}_{-0.0083}$&
$0.0940^{+0.0092}_{-0.0079}$&$0.095^{+0.013}_{-0.010}$\\
pomeron coupling (mb)	&$18.4^{+1.8}_{-2.0}$&
$18.8^{+1.7}_{-2.0}$&$18.5^{+2.1}_{-2.6}$\\\hline
$\rho/\omega$ intercept--1	&
$-0.535^{+0.051}_{-0.059}$&$-0.518^{+0.050}_{-0.058}$
&$-0.540^{+0.059}_{-0.067}$\\
$\rho$/$\omega$ coupling $C_-(pp)$ (mb)& $-31.6^{+7.6}_{-11.5}$ &
$-28.9^{+6.8}_{-10.4}$&$-32.5^{+8.8}_{-13.9}$\\\hline
$a/f$ intercept--1
&$-0.338^{+0.054}_{-0.055}$&$-0.355^{+0.056}_{-0.057}$
&$-0.346^{+0.067}_{-0.066}$\\
$a/f$ coupling (mb)		&
$58.8^{+8.7}_{-6.8}$&$61.5^{+9.8}_{-7.7}$&$60.4^{+10.5}_{-7.9}$\\\hline
$\sigma_{tot}(1.8$ TeV$)$ (mb)	& $77.3^{+2.6}_{-2.7}$&$77.2\pm
2.6$&$77.2^{+3.6}_{-3.3}$\\\hline
$\sigma_{tot}(10$ TeV$)$ (mb) 	&
$106.8^{+7.1}_{-6.3}$&$106.3^{+6.8}_{-6.2}$&$106.5^{+9.7}_{-7.8}$\\\hline
$\sigma_{tot}(14$ TeV$)$ (mb)	&
$113.9^{+8.3}_{-7.4}$&$113.2^{+8.0}_{-7.1}$&$113.5^{+11.4}_{-9.0}$\\\hline
\end{tabular}}~\\
\end{center}\begin{quote}
{\small Table 5: Simple pole fit to total $pp$ and $\bar p p$ cross sections,
and to the $\rho$
parameter, with non-degenerate $C=+1$ and $C=-1$ meson exchanges, and using the
alternative data set of Ref.~\cite{Kangdata}.}
\end{quote}

In order to understand better the treatment of the errors, we give in Table~5
the result of a fit to the data of Ref.~\cite{Kangdata}, where the errors
have been added in quadrature. We see that the results are very stable,
especially those for the pomeron intercept. This is similar to the statement
that
the filtering of the data does not affect the parameters very much.

At this point, the only additional piece of data might be the direct
observation
of the pomeron, {\it i.e.} of a $2^{++}$ glueball.
The WA91 collaboration has indeed confirmed \cite{WA}
 the WA71 observation of the X(1900)
and showed
that it was a $I^GJ^{PC}=0^+2^{++}$ state, $f_2(1900)$.  Its mass has been
observed to be $1918\pm12$ MeV. If we assume that this is the first state
on the pomeron trajectory, and use the determination $\alpha'=0.250$ GeV$^{-2}$
\cite{DL}, we obtain
$\alpha_P=1.0803\pm 0.012$.
This is the value of the intercept for 1-pomeron exchange. The intercepts that
we obtained in Tables~1,~3,~4 and 5 from scattering data
cannot be directly compared
with this value, as they include the effect of multiple exchanges,
of pomerons and reggeons. But the values we have derived are certainly
compatible with the WA91 measurement. Note however that the conversion of the
glueball
mass into a pomeron intercept relies heavily on the value of the pomeron slope.
An intercept of 1.094 would be in perfect agreement with the WA91 observation
for a slope $\alpha'=0.246$ GeV$^{-2}$. Hence it would be dangerous to mix this
piece of
information with the $t$-channel information. Our best estimate for the pomeron
intercept is then:
\begin{equation}
\alpha_P=1.0964^{+0.0115}_{-0.0091}
\label{result}
\end{equation}
based on the $2\sigma$-filtered PDG data, in the non-degenerate case.

Finally, we can place constraints on physics beyond one-pomeron exchange.
The first obvious correction at high energy has to be related to unitarisation.
Clearly, multiple exchanges must tame
the rise of total cross sections so that they eventually obey the
Froissart-Martin bound. Hence one would expect the value of the intercept to
be an effective one, which decreases as $\sqrt{s}$ increases. We have already
seen in Fig.~(1.b) that the data does not show any sign of unitarisation
up to Tevatron energies.
One can confirm this by introducing a 2-pomeron exchange term in the amplitude.
Although little is known about such a contribution, there is general agreement
that it must be negative, and have an intercept $2\alpha_P-1$.
We can then fit the data to a form:
\begin{equation}
Im{\cal A}(s,0)=C_- s^{\alpha_-} + C_+ s^{\alpha_+}
+C_P \left(s^{\alpha_P}+R\ s^{\alpha_{2P}}\right)\label{2pom}
\end{equation}
with $\alpha_{2P}=2\alpha_P-1$.
Using $2\sigma$-filtered data, we then obtain an upper bound on
the ratio $R$ of its coupling to that
of the pomeron:
\begin{equation}
{\rm |2-pomeron\ coupling|\over 1-pomeron\ coupling}<4.7\% \ \ \ \ (70\% C.L.)
\end{equation}
Including such a contribution would bring the best value for the intercept of
the
1-pomeron exchange term to $1.126^{+0.051}_{-0.082}$.

By following the method of Eq.~(\ref{2pom}), one can obtain bounds
on more exotic objects. At the 70\% C.L.,
the ratio of the coupling of an odderon to that of a pomeron is smaller
than 0.1\% (the best odderon intercept would then be 1.105 and the pomeron
intercept become 1.099). This would correspond to 0.08 mb at the Tevatron.
As for the ``hard pomeron'', there is no trace of it in the data.
Constraining
its intercept to being larger than 1.3 leads to an upper bound on the ratio
of its coupling
to that of the pomeron of 0.9\% (the soft pomeron intercept then becomes
1.065). This would correspond to a maximum hard contribution of
 19 mb at the Tevatron.

In conclusion, we have shown that simple pole fits to total cross sections
are very successful.
We show in Fig.~5 the results obtained in this
paper together with other estimates present in the literature.

{}~\\
\centerline{
\psfig{figure=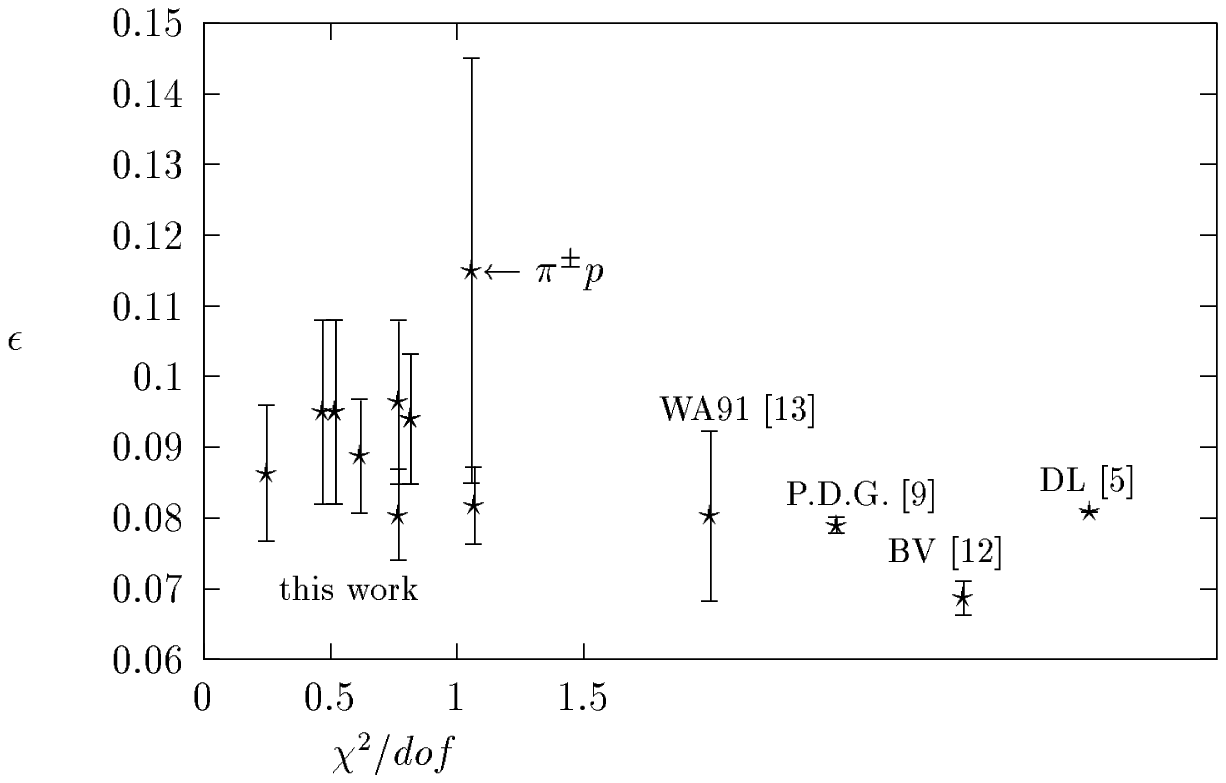,bbllx=2.5cm,bblly=15cm,bburx=17cm,bbury=23cm,width=12cm}
}
\begin{quote}
{\small Figure 5: Our results for the pomeron intercept, compared with others
in the literature. The values of the $\chi^2/dof$ are indicated for the points
of this work only.}
\end{quote}

All the
points from this work  have an acceptable $\chi^2$, and the main difference
between them is either the filtering of data or the physics of lower
trajectories. Hence all these estimates are acceptable, and we feel that
intercepts as high as 1.11, and as low as 1.07, are possible.
When comparing with
other work in the literature, we have explained that the use of a small
energy cutoff leads to smaller intercepts, and reflects
the fact that sub-leading meson trajectories are to be included [note
that the original DL fit \cite{DLfit} used the same cutoff as ours,
but used a different definition of $\chi^2$ \cite{peter}]. Our
errors are much larger than those of other estimates because we fully
take into account the correlation of the various parameters, and because
our statistical analysis of the data is much more careful than previous ones.
In other
words, we quote the projection of the hypersurface containing 70\% of the
probability, rather than letting the $\chi^2$ simply vary by one unit. For
this problem, we believe that this leads to much more reasonable error
estimates, as explained above.

Note that these results depend only on $pp$ and $\bar p p$ data. We have
argued that little could be learned from other hadronic reactions, given
that they are measured at low energy. In particular, we want to point out that
our
fit to total cross sections of Fig.~3 is indistinguishable from the DL fit
for $\sqrt{s}<300$ GeV, hence the parametrisation we propose will fit the
total $\gamma p$ cross section, as well as the $\pi p$ and $K p$ data.
{}~\\
{}~\\

\noindent{\Large \bf Acknowledgments}\\
 We thank P.V. Landshoff for discussion and for sending us
the data set used in the original DL fit. We are also grateful to P. Valin
for his explanation of statistical subtleties. We also wish to acknowledge
discussions with A.~Burnel and P. Marage.

\end{document}